# Estimating demographic parameters using a combination of known-fate and open N-mixture models

JOSHUA H. SCHMIDT, Central Alaska Network, National Park Service, 4175 Geist Road, Fairbanks, Alaska, 99709, USA

DEVIN S. JOHNSON, National Marine Mammal Laboratory, Alaska Fisheries Science Center, National Marine Fisheries Service, NOAA, 7600 Sand Point Way NE, Seattle, Washington 98115, USA

MARK S. LINDBERG, Department of Wildlife and Institute of Arctic Biology, University of Alaska Fairbanks, Fairbanks, Alaska 99775, USA

LAYNE G. ADAMS, United States Geological Survey, Alaska Science Center, 4210 University Drive, Anchorage, Alaska 99508, USA

**Abstract**

1. Accurate estimates of demographic parameters are required to infer appropriate ecological relationships and inform management actions. Known-fate data from marked individuals are commonly used to estimate survival rates, under the assumption that marked individuals represent the unmarked population. Additional information on unmarked individuals is not generally used because of a lack of individual identification, but these unmarked individuals may be more representative and could increase sample sizes thus reducing bias and variance. Recently developed *N*-mixture models use count data from unmarked individuals to estimate demographic parameters, but a joint approach combining the strengths of both analytical tools has not been developed.

2. We present an integrated model combining known-fate and open *N*-mixture models, allowing the estimation of detection probability, recruitment, and the joint estimation of survival. We first use a simulation study to evaluate the performance of the model relative to known values. We also demonstrate how the approach can be used to assess bias in the marked sample relative to the unmarked sample. We then provide an applied example using 4 years of wolf survival data consisting of relocations of radio-collared wolves within packs and counts of associated pack-mates. The model is implemented in both maximum-likelihood and Bayesian frameworks using a new R package kfdnm and the BUGS language.

3. The simulation results indicated that the integrated model was able to reliably recover parameters with no evidence of bias, and estimates were more precise under the joint model as expected. Results from the applied example indicated that the marked sample of wolves was biased towards individuals with higher apparent survival rates (including losses due to mortality and emigration) than the unmarked pack-mates, suggesting estimates of apparent survival based on joint estimation could be more representative of the overall population. Estimates of recruitment were similar to direct observations of pup production, and overlap of the credible intervals suggested no clear differences in recruitment rates.

4. Our integrated model is a practical approach for increasing the amount of information gained from future and existing radio-telemetry and other similar mark-resight datasets. Marking animals is often the most costly aspect of a field project, and our approach could be used to decrease costs, increase precision, and reduce bias inherent in many projects relying on a marked subsample of the population of interest.

**Introduction**
Population ecologists and managers require unbiased and precise estimates of demographic parameters to ensure proper inference (Skalski, Ryding & Millspaugh 2005). Mark-recapture methods are commonly used to estimate survival and other parameters (Williams, Nichols, & Conroy 2002), with radio-marks being particularly useful because marked individuals can be relocated and fate (i.e., alive or dead) can be identified with near certainty (White & Garrott 1990). These field techniques are commonly used to collect survival data in a variety of species including: waterfowl (Ringleman & Longcore 1982; Flint & Grand 1997; Schmidt, Taylor & Rexstad 2006), ungulates (Adams, Singer, & Dale 1995; Johnson et al. 2010; Hebblewhite & Merrill 2011), and carnivores (Adams et al. 2008; Gude et al. 2012). Known-fate models (Pollock, Winterstein, & Conroy 1989) are typically used for analysis of this data type and have a long history of development and application in the ecological literature (e.g., Kaplan & Meier 1958; Trent & Rongstad 1974; Heisey & Fuller 1985; Pollock et al. 1989). In the simplest case, data are limited to the marked individuals, which are assumed to be representative of the population but may also include group members when detection can be assumed to be 1.0 (e.g., ducklings associated with marked hens). In many cases, additional information (e.g., counts, spatial locations) for unmarked individuals are gathered during relocations of the marked sample, but these data are generally not used for estimating demographic rates (although see Johnson et al. 2010) because analytical methods for data from unmarked individuals were previously unavailable.

The development of *N*-mixture models (Royle 2004) has recently provided a method for extracting abundance information from count data that are considered to be of lower value, when compared with mark-recapture data, due to difficulties in interpretation and lack of individual identification. *N*-mixture models allow the estimation of abundance from repeated counts of unmarked individuals by conditioning observed counts on detection probability and abundance, assuming the population is closed (Royle 2004; Chandler, Royle & King 2011; Schmidt, McIntyre & MacCluskie 2013). Abundance may then be estimated for each sample location or at all sites combined by summing the site-level estimates. More recently, open versions of the *N*-mixture model have also been developed (Dail & Madsen 2011; Zipkin et al. 2014b), relaxing the closure assumption and allowing the estimation of population dynamics parameters. This is achieved by assuming that the abundance, $N_{it}$, at each site *i* and time *t* has the Markov property so that $N_{it}$ is dependent only on $N_{it-1}$. Abundance is then modeled as the density of the sum of the number of individuals that survived at site *i* from time *t-1* to time *t*, and the number of individuals that were recruited at site *i* from time *t-1* to time *t* (Dail & Madsen 2011). In the basic form of the model, survival and recruitment are confounded with immigration and emigration, respectively, because of insufficient data. The open *N*-mixture model does not require any closed periods for model identifiability, although the robust design could be incorporated to increase both accuracy and precision (Dail & Madsen 2011). While mark-recapture and *N*-mixture methods are available for estimating demographic parameters using different datasets, separate analyses for each set of data is inefficient.

A natural progression is the integration of a known-fate model based on relocations of radio-marked individuals with an open *N*-mixture model using counts of unmarked group members associated with marked individuals observed during radio relocations. The combination of data from multiple sources to improve inference has received much attention, largely due to the prospects for increasing precision, reducing required sample sizes, and estimating additional parameters (e.g., Borchers 2012; Sollmann et al. 2013). Integrated population models (Besbeas et al. 2002; Schaub et al. 2007; Abadi et al. 2010; Schaub & Abadi 2012; Chandler & Clark 2014) incorporate multiple data sources to jointly estimate demographic parameters. Integrated models are likely to be more efficient at large spatial scales (Zipkin et al. 2014a) and may be particularly useful for telemetry studies, which tend to be expensive both logistically and monetarily. For social species, most of the observed individuals are unmarked, but due to frequent relocation of radio-marked group members, much additional information on survival and recruitment in the form of repeated counts could be used. To our knowledge, an integrated model combining known-fate and open *N*-mixture models has not been developed. A combined approach could increase the amount of demographic information for a variety of past and future studies.

Here we present an approach combining known-fate and open *N*-mixture models to jointly estimate the survival parameter while also providing estimates of annual recruitment. This model uses capture histories from radio-collared individuals and associated counts of unmarked individuals observed with those that are marked. We expected that combining these two data types would result in more accurate and precise estimates, particularly of survival, and would allow the estimation of recruitment, which is often not possible using only the information from radio-marked individuals. We use simulations to assess the ability of our integrated model to recover parameters and demonstrate a simple approach for assessing bias in the marked sample. We also provide a practical example by applying our joint model to 4 years of previously published wolf (*Canis lupus*) survival and pack count data collected by relocating a subset of radio-marked individuals within packs (Adams et al. 2008). A general comparison of results with those of Adams et al. (2008) illustrates the similarities and differences in inference that may be possible with an integrated approach. We provide implementations in both the BUGS language and an R package.

MODEL DESCRIPTION

To share strength between the known-fate and open *N*-mixture data types, we combined a known-fate survival submodel and an open *N*-mixture submodel in a hierarchical fashion to estimate demographic parameters with increased precision. First, for the known-fate submodel we used the same basic structure as previously published hierarchical nest survival models (Royle & Dorazio 2008; Schmidt et al. 2010). Because fates are assumed to be known with certainty for known-fate individuals, detection probability is not estimated. Although many known-fate models are formulated at the individual level (e.g., Pollock, Winterstein, & Conroy 1989), here we assume there are negligible individual effects within known groups thus, survival is constant between individuals within groups. However, shared covariates could be used to relax this assumption. We model the status of the *m*th known-fate individual of group *i*, at time *t*, $Y_{mit}$, as

$$[Y_{mit}|Y_{mi,t-1}] = \text{Bernoulli}(\omega_{i,t-1} \cdot Y_{mi,t-1}); \; m = 1, \ldots, M, i = 1, \ldots, I, \text{and } t = 2, \ldots, T,$$

where $[x|y]$ represents the conditional probability distribution of $x$ given $y$, $\omega_{i,t-1}$ is the probability of survival from time *t* -1 to *t*. The survival probabilities are modeled using

$$\text{logit } \omega_{i,t-1} = \mathbf{x}'_{i,t-1}\boldsymbol{\beta}$$

where $\mathbf{x}_{i,t-1}$ is a vector of known covariates and $\boldsymbol{\beta}$ are the associated coefficients.

In addition to the known-fate individuals in group *i* at time *t*, we also assume there are $N_{it}$ additional individuals in the group. These additional individuals are composed of those who survived from the previous time, $S_{it}$, to the current time plus those that are recruited to the group at the current time, $G_{it}$. It may become necessary to replenish the sample of known-fate individuals due to accumulating deaths, so, the known-fate sample does not become extinct through the course of the study. Therefore, let $R_{it}$ denote the known number of individuals removed from the general population at time $t$ and placed into the known-fate sample.

The survivors are modeled via,
$$[S_{it}|N_{i,t-1}] = \text{Binomial}(N_{i,t-1} - R_{i,t-1}, \omega^*_{i,t-1}),$$
where
$$\text{logit } \omega^*_{i,t-1} = \mathbf{x}'_{i,t-1}\boldsymbol{\beta}^*.$$

To share information between the two data sets and see improvements to parameter inference we assume that some, if not all, elements of $\boldsymbol{\beta}$ are equal to $\boldsymbol{\beta}^*$. Following Dail and Madsen (2011), the recruited individuals are modeled with,
$$[G_{it}] = \text{Poisson}(\gamma_{it}),$$
where $\gamma_{it}$ is the recruitment rate parameter. To allow inclusion of covariate information, $\mathbf{w}_{it}$, we can further parameterize the recruitment rate using,
$$\log \gamma_{it} = \mathbf{w}'_{it}\boldsymbol{\rho}.$$

Finally, as with all open $N$-mixture data, the general population is usually not observed, therefore a detection model is necessary to model the observed abundance, $n_{it}$. Given the true abundance of the general population, we assume the observation process represents a binomial sample, i.e.,
$$[n_{it}|N_{it}] = \text{Binomial}(N_{it} - R_{it}, p_{it}),$$
where $p_{it}$ is the probability of detecting one of the members of the group $i$ in the general population at time $t$, and $R_{it}$ are the number removed from the general population to the known-fate sample. Here we formulate the model with removals counted before the detection process. In our opinion this makes more sense as the process of sampling new known-fate individuals is probably more involved and invasive that the usual survey methodology for the open $N$-mixture detection process. Therefore, we do not want to model those individuals with a $p_{gt}$ detection probability. Analogous to the survival models, we parameterize the detection probability using,
$$\text{logit } p_{it} = \mathbf{z}'_{it}\boldsymbol{\alpha}$$
where $\mathbf{z}_{it}$ is a vector of known covariates and $\boldsymbol{\alpha}$ are the associated coefficients.

Using the dynamic components of the model we can derive the transition kernel of the abundance process using the appropriate convolution (Dail and Madsen, 2011),
$$[N_{it} = k|N_{i,t-1} = j] = \sum_{c=0}^{\min\{j-R_{i,t-1},k\}} \text{Binomial}(c|j - R_{i,t-1}, \omega^*_{i,t-1})\,\text{Poisson}(k - c|\gamma_{it}),$$
Note, our definition is slightly altered to account for the known removals, $R_{it}$, to the known-fate sample.

Known-fate open $N$-mixture inference

Dail and Madsen (2011) first proposed a maximum likelihood approach for making inference for open $N$-mixture model parameters by using the transition kernel for the $N$ process. Here, we augment their full-data likelihood with the known-fate portion of the model to obtain the known-fate-open $N$-mixture full-data likelihood
$$[\boldsymbol{\beta}, \boldsymbol{\beta}^*, \boldsymbol{\rho}, \boldsymbol{\alpha}|\mathbf{n}, \mathbf{R}, \mathbf{N}] = \prod_{m=1}^{M}\prod_{i=1}^{I}\prod_{t=1}^{T}[Y_{mit}|Y_{mi,t-1}] \times \prod_{i=1}^{I}\prod_{t=1}^{T}[n_{it}|N_{it}]\,[N_{it}|N_{i,t-1}]$$
where the bold vectors represent complete collections of the associated data and abundance, $[N_{i1}|N_{i0}]=[N_{i1}]$ represents the prior or initial distribution of abundance at the first time period.

Dail and Madsen (2011) use a Poisson distribution, however, we propose the scale prior $[N_{i1}]=1/N_{i1}$ as a non-informative alternative (Link 2013). Note that the full data likelihood is separable into the known-fate and open *N*-mixture model portions. The known-fate portion is simply a product of conditional Bernoulli distributions, thus, that portion of the likelihood is readily computed. The full-data portion of the open *N*-mixture likelihood must be integrated over the latent abundance process to obtain the true likelihood of that portion of the model.

Here we focus on an efficient method to calculate the true likelihood of the open *N*-mixture model integrated over the dynamic abundance processes. Using the Markov transition kernel of the abundance process, $[N_{it}|N_{i,t-1}]$, the open *N*-mixture model can be formulated as a Hidden Markov Model (HMM; see Zucchini and MacDonald 2009) from which the log-likelihood is efficiently calculated using the forward algorithm (Zucchini and MacDonald 2009; pg. 47). When combined with the backward sampling algorithm, an efficient MCMC algorithm can be devised for Bayesian inference.

In the description, we will provide the definition of the HMM forward algorithm for just a single group. The total likelihood can then be calculated by summing the individual log-likelihoods. First, as in Dail and Madsen (2011), let *K* be the defined upper bound for all $N_{it}$. We assume that for most, if not all, applications, this can be chosen appropriately. Then, let $\boldsymbol{\eta}$ be the row vector of initial abundance probabilities, $[N_1 = j]; j = 0, \ldots, K$ from the chosen initial distribution. Next define $\mathbf{P}(n_{it})$ to be the diagonal matrix with entries Binomial($n_{it}|j - R_{it}, p_{it}$); $j = R_{it}, \ldots, K$ and 0 for $j < R_{it}$. Finally, $\boldsymbol{\Delta}_{t-1}$ is the state transition matrix with *j,k* entry $[N_{it} = k|N_{i,t-1} = j]$. Now, the HMM forward algorithm for calculation of the log-likelihood, *l*, proceeds as follows:

(1) Set:
$w_1 = \boldsymbol{\eta}\mathbf{P}(n_{i1})\mathbf{1}, \boldsymbol{\phi}_1 = \boldsymbol{\eta}\mathbf{P}(n_{i1})/w_1$, and $l = \log w_1$,
(2) For *t* = 2, ..., *T*:
$w_t = \boldsymbol{\phi}_{t-1}\boldsymbol{\Delta}_{t-1}\mathbf{P}(n_{it})\mathbf{1}, \boldsymbol{\phi}_t = \boldsymbol{\phi}_{t-1}\boldsymbol{\Delta}_{t-1}\mathbf{P}(n_{it})/w_t$, and $l = l + \log w_t$.

While the reverse-time recursion of Dail and Madsen (2011) provides a computationally efficient method to calculate the open *N*-mixture models likelihood, the HMM also provides efficient methods for Bayesian inference via MCMC. An outline of an MCMC routine proceeds as follows:

(1) For current parameter vector, $\boldsymbol{\theta}^{curr} = (\boldsymbol{\beta}, \boldsymbol{\beta}^*, \boldsymbol{\rho}, \boldsymbol{\alpha})$,
   1. Draw, $\boldsymbol{\theta}^{prop}$, from proposal distribution $[\boldsymbol{\theta}^{prop}|\boldsymbol{\theta}^{curr}]$
   2. Set $\boldsymbol{\theta}^{curr} = \boldsymbol{\theta}^{prop}$ with probability

$$\min\left(1, \exp\{l(\boldsymbol{\theta}^{prop}) - l(\boldsymbol{\theta}^{curr})\} \frac{[\boldsymbol{\theta}^{prop}] \cdot [\boldsymbol{\theta}^{curr}|\boldsymbol{\theta}^{prop}]}{[\boldsymbol{\theta}^{curr}] \cdot [\boldsymbol{\theta}^{prop}|\boldsymbol{\theta}^{curr}]}\right),$$

   where $l(\boldsymbol{\theta})$ is the log-likelihood evaluated at $\boldsymbol{\theta}$, calculated using the forward algorithm and $[\boldsymbol{\theta}]$ is the prior distribution of the parameters.

If an MCMC sample of the abundance vector, **N**, is desired proceed to (2), otherwise repeat step one as desired to obtain a sufficient posterior sample.

(2) Given a value, $\boldsymbol{\theta}$, from (1):
   1. Run the forward algorithm and retain the $\boldsymbol{\phi}_t$ and $\boldsymbol{\Delta}_t$. The forward run from (1) can be used to avoid re-running the algorithm.
   2. Draw from $[N_{gT}|\mathbf{n}] = \boldsymbol{\phi}_T$.

3. For t = *T*-1,…,1:

   Draw from $[N_{it}|N_{i,t+1}, \ldots, N_{iT}, \boldsymbol{\theta}, \mathbf{n}] \propto \boldsymbol{\phi}_t \cdot \boldsymbol{\delta}_t^{N_{i,t+1}}$, where the product is element-wise and $\boldsymbol{\delta}_t^{N_{i,t+1}}$ is the $N_{i,t+1}$ column $\boldsymbol{\Delta}_t$ (Zucchini and MacDonald 2009).

(3) If desired, the $S_{it}$ and $G_{it}$ processes can be directly sampled following updates of the $N_{it}$ process:
   1. Draw from:
      $[S_{it}|N_{i,t-1}, N_{it}] \propto \text{Binomial}(S_{it}|N_{i,t-1} - R_{i,t-1}, \omega^*)\text{Poisson}(N_{i,t} - S_{it}|\gamma_{it})$,
      for $S_{it}$ in $\{0,\ldots, N_{it} - R_{it}\}$,
   2. Set $G_{it} = N_{it} - S_{it}$.

By sampling the hidden state, **N**, as a single vector versus individually given immediate neighbors, i.e., not drawing from $[N_{it}|N_{i,t-1}, N_{i,t+1}]$, with in the MCMC, high autocorrelation and a slowly converging chain can be avoided. In addition, sampling of the $N_{it}$ process by construction via serially correlated $S_{it}$ and $G_{it}$ samples can be avoided. Of course, if direct inference on the $N_{it}$ process itself is not desired, step (2) is not necessary for Bayesian inference of $\boldsymbol{\theta}$. This reduces Monte Carlo autocorrelation in $\boldsymbol{\theta}$ due to parameter updates being conditioned on the latent abundance updates.

We present both maximum-likelihood and Bayesian implementations of our model using programs R 3.1.0 (R Core Development Team 2014) and OpenBUGS 3.2.3 (Thomas et al. 2006). We also created an R package kfdnm[1] containing both maximum-likelihood and MCMC implementations using the HMM formulation. Code for a Bayesian implementation of the model in OpenBUGS can be found in Appendix S1.

SIMULATIONS

Using both the kfdnm package and OpenBUGS, we ran two sets of simulations to assess the ability of our model to jointly estimate both survival and recruitment. We considered scenarios for a 5 year project with 10 revisits per year with the assumption that three individuals in each of 20 groups carried radio marks at the beginning of each year. The initial number of unmarked individuals present in each group was drawn form a Poisson distribution with a mean of 4.5. At each revisit, the fate of the marked individuals was observed with certainty and the true count of the unmarked group-mates was partially observed, assuming the probability of detection was p = 0.5. For the purposes of this study, we assumed that immigration and emigration did not occur, although this assumption could be relaxed if appropriate data were available. The number of recruits in each group was assumed to come from a Poisson distribution with a mean of 4.0, and the recruitment event was assumed to occur in the first time period each year. First we considered a scenario designed to separately estimate survival for both marked ($\phi_k^1 = 0.90$) and unmarked ($\phi_k^2 = 0.90$) individuals (i.e., $\phi_k$ not shared). This is analogous to fitting a known-fate model to data from the marked subset and a separate open *N*-mixture model to the repeated count data, thereby providing a test of the assumption that the marked sample is representative of the population of interest. We also simulated data where $\phi_k = 0.90$ for all individuals and survival was jointly estimated (i.e., $\phi_k$ was shared). We generated 200 datasets for each scenario and fit our model to each replicate set, saving the mean of the posterior of each parameter of interest to assess the ability of the model to recover the data-generating values.

---

[1] The package is available at https://github.com/NMML/kfdnm/releases

The results of our simulations showed that the implementations in OpenBUGS and **kfdnm** recovered the data-generating parameter estimates under both scenarios considered (Appendix S2). The accuracy of the estimates when survival was estimated separately indicates that using the model to test the assumption of representativeness of the marked sample is valid. In addition, when the survival parameter is estimated jointly using both data sources, both survival and recruitment estimates are accurate and more precise than when estimated separately. Attempts to fit our model in WinBUGS (Lunn et al. 2000) and JAGS (Plummer 2003) were unsuccessful (results not shown).

APPLICATION: WOLF RADIO-TELEMETRY DATA
The wolf population in Gates of the Arctic National Park and Preserve (GAAR), Alaska, was studied from the spring of 1987 to spring 1991 to investigate population dynamics and the effects of human harvest (see Adams et al. 2008). We applied our model to the GAAR wolf data as an example with the goals of identifying potential bias in the marked sample and providing a general comparison of survival and recruitment estimates to those based on known-fate methods and direct observations. Radio-marked wolves were maintained in 14-19 packs between April 1987 and January 1990, declining to 8 monitored packs by April 1991. Packs were relocated and individuals were counted throughout the year, although effort varied throughout the study. In the original study, loss of wolves from the marked population was separated into mortality versus emigration, and annual estimates were calculated for each using known-fate methods (Heisey & Fuller 1985). The maximum number of pups observed per pack in September-October on average each year was used as an estimate of annual recruitment. Final survival estimates were based on data pooled across age, sex, and years, largely due to limited sample sizes (see Adams et al. 2008 for further details).

We did not distinguish between losses due to emigration versus mortality, so our estimates of survival are interpreted as apparent survival and include both types of losses of individuals from packs. During many months, radio-marked wolves were often relocated multiple times, although we consolidated the data to a single observation for each month using the highest observed count as the best observation of the unmarked individuals. We assumed over counting did not occur. If a radio-marked wolf was no longer able to be relocated due to a lack of signal, it was right censored; Adams et al. (2008) treated these as dispersers. Not all packs were located during each month, resulting in many missing values. When multiple records of the same number of individuals within a pack were recorded during a single month under good sighting conditions, we assumed $p = 1.0$ for that time period. While 'perfect' counts are not required, their inclusion improved estimation. Otherwise, we assumed detection probability was constant across months. We also assumed survival was constant across months and years, but allowed recruitment to vary by year. Recruitment, defined as additions of individuals to a pack, was assumed to occur during May when pups are typically born. Therefore, recruitment only occurred during the first month of the biological year (May 1 of the current year through April 30 of the following year). We assumed that immigration did not occur, however, if immigrants were added to some packs, estimates of recruitment would likely be biased high.

We fit two versions of the model to the GAAR data using both OpenBUGS and **kfdnm**. We estimated survival separately for the marked and unmarked individuals in the first version of the model, while in the second we jointly estimated survival. While there were pup observations available for many packs during June-October, when pups were out of their dens and distinguishable from adults, we did not include this information in the dataset in order to see how well the resulting estimates matched the observed recruitment values as reported by Adams et al. (2008). Further extensions could also incorporate counts of young, when available, to aid estimation of the recruitment parameter.

We found that annual survival rates were lower in the unmarked sample than in the marked sample when estimated separately for the two groups (Fig. 1A). Estimates for the marked sample alone were very similar to those from Adams et al. (2008) as expected (Fig. 1A). Results from the joint model showed that survival for the sampled population (marked and unmarked) was likely lower than that based on the collared sample alone (Fig. 1A). The estimated numbers of individuals added during May were generally larger than observed average pup counts and were less precise in later years, corresponding to reduced numbers of relocations and reduced numbers of packs in the sample (Fig. 1B). Similar to the observed pup counts, our estimates exhibited an increasing trend, but the 95% credible intervals around these estimates overlapped for all 4 years, indicating no clear differences in recruitment among years. Together our results suggest that apparent survival was lower in the overall population as compared to the marked population alone and that any trend in estimated recruitment was unclear. Results produced using kfdnm and OpenBUGS were very similar (Fig. 1).

We also fit a model using a spline function allowing survival to vary across months in the same pattern between collared and uncollared animals, while the 2 groups were assumed to have different overall survival rates. This allowed sharing of information on the annual pattern of survival among groups, despite the assumption that annual rates differed. We fit this model using both MLE and MCMC methods using the kfdnm package to explore variation in survival throughout the year and compare estimates of precision between the two formulations. These results suggested that monthly survival rates declined until late winter before increasing into spring (Fig. 2). They also showed that the estimates from the Bayesian implementation were more precise. Other covariates could be used in a similar manner to share information between groups and improve estimation when bias in the marked sample is suspected.

**Discussion**

Through simulations and a practical example, we demonstrated that by combining known-fate data from radio-marked individuals with count data from associated group-mates, unbiased estimates of survival and recruitment can be produced that are more accurate and precise than is possible using typical known-fate approaches. Combining a standard known-fate model with an open *N*-mixture model allows the joint estimation of survival, in addition to recruitment. Many researchers employing radio-tags collect associated count data during relocations of marked animals. Here we have shown how these additional data can be directly included in the analysis to reduce bias and increase precision, indicating that auxiliary data should generally be collected when sampling marked individuals (Pollock 2002; Lindberg 2012). We also provided a straightforward approach for assessing bias in the marked sample by separately estimating survival for the marked and unmarked samples. Even if such bias exists, the inclusion of covariates can be used to share information among groups to improve estimation. In addition, the ability to directly estimate recruitment is appealing and could further increase the amount of information gained from these studies. Overall, we expect many projects utilizing known-fate methods to assess survival rates in group-dwelling animals would benefit from our integrated analytical approach.

The results of our simulations demonstrated that our model performed accurately. Under each scenario the model reliably recovered the true parameter values, and while it may not be surprising that precision increased when survival was estimated jointly, these results confirm that increases in precision are possible when count data are also utilized. Interestingly, when attempting to use either the WinBUGS or JAGS software packages, estimates for several parameters were consistently biased relative to the generated values for reasons that were not apparent. Identical model code showed no evidence of bias when fit using OpenBUGS or the kfdnm R package. This finding is consistent with that of Kery and Schaub (2012:410) who reported that efforts to fit the open *N*-mixture model in WinBUGS were unsuccessful. The patterns of bias we observed were similar to those found by Zipkin et al. (2014b) for a stage-structured open *N*-mixture model, possibly providing an alternate explanation for the skewed estimates they observed. We suspect that differences in the selection of algorithms between the different software packages may explain the inconsistent performance of open *N*-mixture approaches in different implementations of the BUGS language. Although further work will be required to determine the cause of the bias in other software packages, we thought the potential for unexplained bias may be of interest to others working with similar models.

While our estimates of survival are not directly comparable to those of Adams et al. (2008) our apparent survival estimates for the marked sample alone were similar and consistent with differences in application of the available data (e.g., treatment of emigrants), thus suggesting our results overall are analogous. We did find that the magnitude of the bias in estimated apparent survival rates between the marked and unmarked sample was fairly large ($\phi_{annual}^1 = 0.74$ versus $\phi_{annual}^2 = 0.61$). The finding of some sample bias is not surprising because adults that are less likely to be lost from the population through emigration are commonly targeted for capture and marking. Even if selected randomly, older individuals with higher apparent survival tend to accumulate in the marked sample over time as younger marked animals with markedly lower apparent survival are lost and additional radio-marks are deployed amongst the remaining pack members (Adams et al. 2008). Bias induced by capture heterogeneity (Fletcher et al. 2012), an aging sample (Prichard, Joly, & Dau 2012), or individual heterogeneity (Vaupel & Yashin 1985; Lindberg, Sedinger, & Lebreton 2013) can have implications for assessing population growth rates. We expect that sample bias and marking effects (Murray & Fuller 2000) could be substantial in many settings, possibly leading to inappropriate conclusions and comparisons among populations. Our model relaxes the assumption of a representative sample of marked individuals and provides inference to the entire population. We found fewer differences in estimated recruitment between our work and that of Adams et al. (2008), although our estimates tended to be higher because they represent the average number of individuals added to each pack in May rather than those surviving until direct observations occurred (June-October).

The combination of multiple data sources to improve estimation is an active area of development and promises to increase the amount of demographic information that can be extracted from commonly collected field data (e.g., Kery & Schaub 2012; Bird et al. 2014). While we have demonstrated how count and known-fate data may be combined to improve inference, the inclusion of other data sources and demographic parameters is also possible. For example, composition data or information on movement rates could be incorporated to help estimate cohort-specific survival and immigration/emigration rates, respectively (e.g., Zipkin et al 2014b). For simplicity, we did not include covariates in our demonstration, although they could be easily added for other applications. Our simple example using a shared spline function demonstrated that the inclusion of covariates could be used to share information among groups. Additionally, implementation in a Bayesian framework provides a mechanism for the inclusion of prior information that may be used to further increase precision (McCarthy & Masters 2005; Schmidt & Rattenbury 2013).

Further extensions of our model could include the incorporation of the robust design if shorter closed periods were sampled between open periods. The stage structured model of Zipkin et al. (2014b) may also benefit from the inclusion of known-fate data when available. With slight modifications, our model could also be useful in quantifying survival as well as immigration and emigration rates in species with more fluid group dynamics (e.g., muskoxen *Ovibos moschatus*), if repeated counts of multiple groups containing marked individuals were available. It is also possible that other models could be used in place of the known-fate model, with similar benefits for other types of studies (e.g., distance sampling; Sollmann et al. 2015). The integration of mark-recapture or distance sampling models with open *N*-mixture models should provide opportunities to improve ecological inference in a variety of settings and will help to increase our knowledge of population dynamics. As a general approach, explicitly combining data sources can provide a much more complete picture of the population dynamics of a species than would be possible through independent analytical efforts.

**Acknowledgements**
Funding for this work was provided by the U.S. National Park Service's Inventory and Monitoring Program through the Central Alaska Network. The findings and conclusions in the paper are those of the authors and do not necessarily represent the views of the National Marine Fisheries Service, NOAA, the U.S. National Park Service, or the U.S. Geological Survey. Reference to trade names is for descriptive purposes only and does not imply endorsement by the U.S. government.

**References**
Abadi, F., Giminez, O., Arlettaz, R., & Schaub, M. (2010) An assessment of integrated population models: bias, accuracy, and violation of the assumption of independence. *Ecology*, 91, 7-14.
Adams, L.G., Singer, F.J., & Dale, B.W. (1995) Caribou calf mortality in Denali National Park, Alaska. *Journal of Wildlife Management*, 59, 584-594.
Adams, L.G., Stephenson, R.O., Dale, B.W., Ahgook, R.T. & Demma, D.J. (2008) Population dynamics and harvest characteristics of wolves in the central Brooks Range, Alaska. *Wildlife Monographs,* 170, 1-25.
Beabeas, P., Freeman, S.N., Morgan, B.J.T., and Catchpole, E.A. (2002) Integrating mark-recapture-recovery and census data to estimate animal abundance and demographic parameters. *Biometrics*, 58, 540-547.
Bird, T., Lyon, J., Nicol, S., McCarthy, M., & Barker, R. (2014) Estimating population size in the presence of temporary migration using a joint analysis of telemetry and capture-recapture data. *Methods in Ecology and Evolution*, 5, 615-625.
Borchers, D. (2012) A non-technical overview of spatially explicit capture-recapture models. *Journal of Ornithology*, 152, S435-S444.
Brooks, S.P., King, R., & Morgan, B.J.T. (2004) A Bayesian approach to combining animal abundance and demographic data. *Animal Biodiversity and Conservation*, 27, 514-529.
Chandler, R.B. & Clark, J.D. (2014) Spatially explicit integrated population models. *Methods in Ecology and Evolution*, DOI: 10.1111/2041-210X.12153.

Chandler, R. B., Royle, J. A., & King, D. I. (2011) Inference about density and temporary emigration in unmarked populations. *Ecology*, 92, 1429-1435.

Dail, D. & Madsen, L. (2011) Models for estimating abundance from repeated counts of an open metapopulation. *Biometrics* 67, 577-587.

Fletcher, D., Lebreton, J.-D., Marescot, L., Schaub, M., Gimenez, O., Dawson, S., & Slooten, E. (2012) Bias in estimation of adult survival and asymptotic population growth rate caused by undetected capture heterogeneity. *Methods in Ecology and Evolution*, 3, 206-216.

Flint, P.L. & Grand, J.B. (1997) Survival of spectacled eider adult females and ducklings during brood rearing. *Journal of Wildlife Management*, 61, 217-221.

Gude, J.A., Mitchell, M.S., Russell, R.E., Sime, C.A., Bangs, E.E., Mech, L.D. & Ream, R.R. (2012) Wolf population dynamics in the U.S. northern Rocky Mountains are affected by recruitment and human-caused mortality. *Journal of Wildlife Management*, 76, 108-118.

Hebblewhite, M. & Merrill, E.H. (2011) Demographic balancing of migrant and resident elk in a partially migratory population through forage-predation tradeoffs. *Oikos*, 120, 1860-1870.

Heisey, D.M., & Fuller, T.K. (1985) Evaluation of survival and cause-specific mortality rates using telemetry data. *Journal of Wildlife Management*, 49, 668-674.

Johnson, H.E., Mills, L.S., Wehausen, J.D. & Stephenson, T.R. (2010) Combining ground count, telemetry, and mark-resight data to infer population dynamics in an endangered species. *Journal of Applied Ecology*, 47, 1083-1093.

Kaplan, E.L., & Meier, P. (1958) Nonparametric estimation from incomplete observations. *Journal of the American Statistical Association*, 53, 457-481.

Kery, M., & Schaub, M. (2012) *Bayesian population analysis using WinBUGS: a hierarchical perspective*. Elsevier Academic Press, San Diego, California, USA.

Lindberg, M.S. (2012) A review of designs for capture–mark–recapture studies in discrete time. *Journal of Ornithology*, 152, 355-370.

Lindberg, M.S., Sedinger, J.S., & Lebreton, J.-D. (2013) Individual heterogeneity in black brant survival and recruitment with implications for harvest dynamics. *Ecology and Evolution*, 3, 4045-4056.

Link, W.A. (2013) A cautionary note on the discrete uniform prior for the binomial *N*. *Ecology*, 94, 2173-2179.

Lunn, D. J., Thomas, A., Best, N., & Spiegelhalter, D. (2000) WinBUGS–a Bayesian modeling framework: concepts, structure, and extensibility. *Statistical Computing*, 10, 325-337.

McCarthy, M.A. & Masters, P. (2005) Profiting from prior information in Bayesian analyses of ecological data. *Journal of Applied Ecology*, 42, 1012-1019.

Murray, D.L., & Fuller, M.R. (2000) A critical review of the effects of marking on the biology of vertebrates. In: Boitani, L., & Fuller, T. (eds) *Research techniques in animal ecology: controversies and consequences*. Columbia University Press, New York, pp 15–64.

Plummer, M. (2003) JAGS: a program for analysis of Bayesian graphical models using Gibbs Sampling. *Proceedings of the Third International Workshop on Distributed Statistical Computing*. R Project for Statistical Computing, Vienna, Austria.

Pollock, K. H. (2002) The use of auxiliary variables in capture-recapture modeling : An overview. *Journal of Applied Statistics*, 29, 85-102.

Pollock, K.H., Winterstein, S.R., Bunck, C.M., & Curtis, P.D. (1989) Survival analysis in telemetry studies: the staggered entry design. *Journal of Wildlife Management*, 53, 7-15.

Pollock, K.H., Winterstein, S.R., & Conroy, M.J. (1989) Estimation and analysis of survival distributions for radio-tagged animals. *Biometrics*, 45, 99-109.
Prichard, A.K., Joly, K., & Dau, J. (2012) Quantifying telemetry collar bias when age is unknown: a simulation study with a long-lived ungulate. *Journal of Wildlife Management*, 76, 1441-1449.
R Development Core Team. (2012) *R: a language and environment for statistical computing*. R Foundation for Statistical Computing, Vienna, Austria.
Ringleman, J.K. & Longcore, J.R. (1982) Survival of juvenile black ducks during brood rearing. *Journal of Wildlife Management*, 46, 622-628.
Royle, J.A. (2004) *N*-mixture models for estimating population size from spatially replicated counts. *Biometrics*, 60, 108-115.
Royle, J.A. & Dorazio, R.M. (2008) *Hierarchical modeling and inference in ecology: the analysis of data from populations, metapopulations, and communities.* Academic Press, New York, USA.
Schaub, M. & Abadi, F. (2012) Integrated population models: a novel analysis framework for deeper insights into population dynamics. *Journal of Ornithology,* 152, S227-S237.
Schaub, M., Gimenez, O., Sierro, A., & Arlettaz, R. (2007) Use of integrated modeling to enhance estimates of population dynamics obtained from limited data. *Conservation Biology*, 21, 945-955.
Schmidt, J.H. & Rattenbury, K.L. (2013) Reducing effort while improving inference: estimating Dall's sheep abundance and composition in small areas. *Journal of Wildlife Management*, 77, 1048-1058.
Schmidt, J.H., Taylor, E.J. & Rexstad, E.A. (2006) Survival of common goldeneye ducklings in interior Alaska. *Journal of Wildlife Management*, 70, 792-798.
Schmidt, J.H., Walker, J.A., Lindberg, M.S., Johnson, D.S. & Stephens, S.E. (2010) A general Bayesian hierarchical model for estimating survival of nests and young. *Auk*, 127, 379-386.
Skalski, J.R., Ryding, K.E., & Millspaugh, J.J. (2005) *Wildlife demography: analysis of sex, age, and count data.* Elsevier Academic Press, San Diego, California, USA.
Sollmann, R., Gardner, B., Chandler, R.B., Royle, J.A., & Sillett, T.S. 2015. An open population hierarchical distance sampling model. *Ecology*, http://dx.doi.org/10.1890/14-1625.1
Sollmann, R., Gardner, B., Chandler, R.B., Shindle, D.B., Onorato, D.P., Royle, J.A., & O'Connell, A.F. (2013) Using multiple data sources provides density estimates for endangered Florida panther. *Journal of Applied Ecology*, 50, 961-968.
Thomas, A., O'Hara, B.O., Ligges, U., & Sturtz, S. (2006) Making BUGS open. *R News*, 6, 12-17.
Trent, T.T., & Rongstad, O.J. (1974) Home range and survival of cottontail rabbits in southwestern Wisconsin. *Journal of Wildlife Management*, 38, 459-472.
Vaupel, J. W., & Yashin, A. I. (1985) Heterogeneity's ruses: some surprising effects of selection on population dynamics. *American Statistician* 39, 176–185.
White, G.C., & Garrott, R.A. (1990) *Analysis of wildlife radio-tracking data*. Academic Press, San Diego, California, USA.
Williams, B.K., Nichols, J.D., & Conroy, M.J. (2002) *Analysis and management of animal populations*. Academic Press, San Diego, California, USA.

Zipkin, E.F., Sillett, T.S., Campbell Grant, E.H., Chandler, R.B. & Royle, J.A. (2014a) Inferences about population dynamics from count data using multistate models: a comparison to capture-recapture approaches. *Ecology and Evolution*, 4, 417-426.

Zipkin, E.F., Thorson, J.T., See, K., Lynch, H.J., Campbell Grant, E.H., Kanno, Y., Chandler, R.B., Letcher, B.H., & Royle, J.A. (2014b) Modeling structured population dynamics using data from unmarked individuals. *Ecology*, 95, 22-29.

Zucchini, W., & MacDonald, I.L. (2009) *Hidden Markov Models for Time Series: An Introduction Using R*. CRC Press, New York. 275 pp.

**Supporting Information**
Additional Supporting Information may be found in the online version of this article.
**Appendix S1.** OpenBUGS code for the combined known-fate open *N*-mixture model.
**Appendix S2.** Results of kfdnm and OpenBUGS simulations

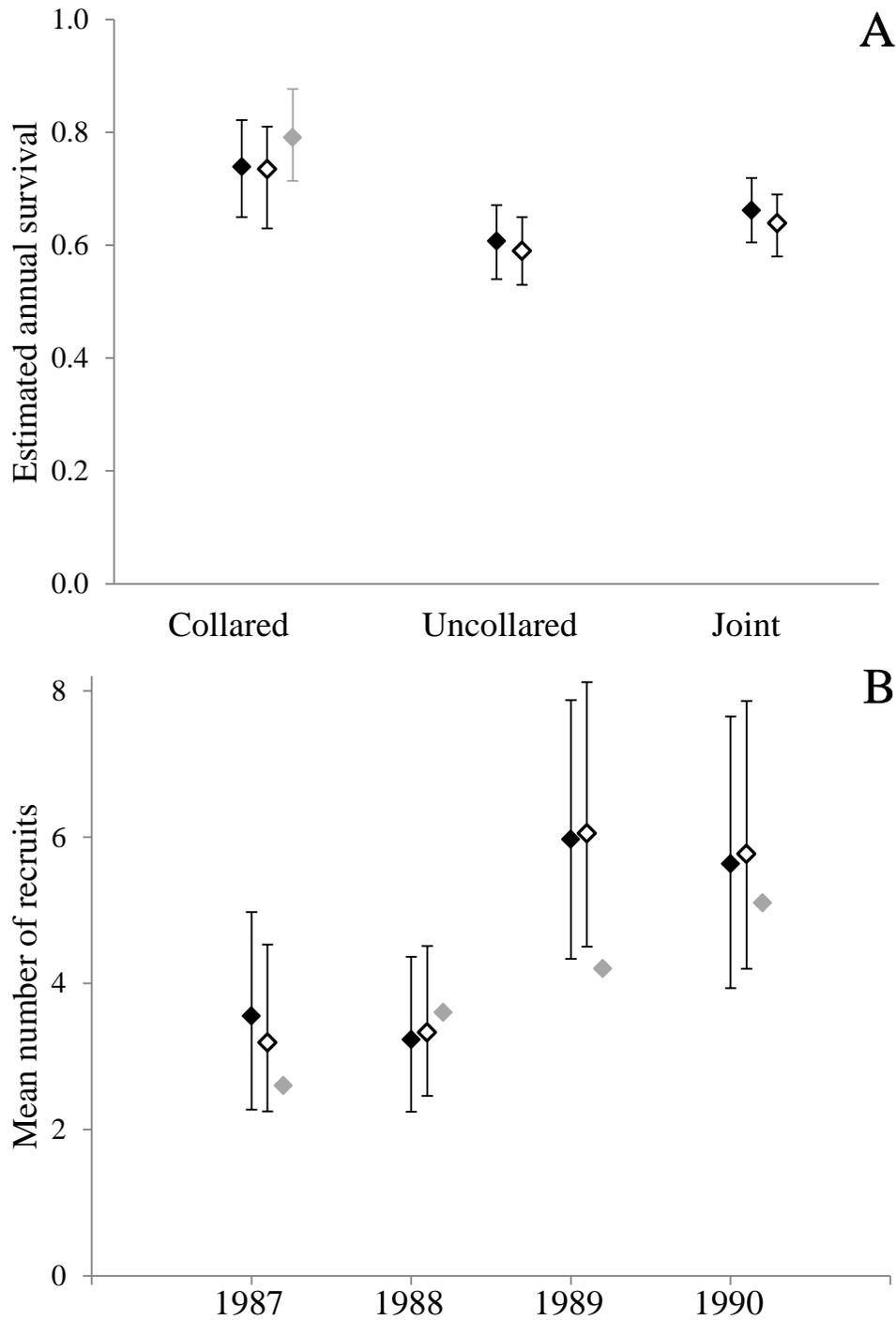

Fig. 1. (A) Estimated mean annual apparent survival probabilities (includes mortality and emigration) for the wolf population studied in Gates of the Arctic National Park and Preserve, Alaska from 1987-1990. Two sets of estimates are shown: survival of collared and uncollared wolves estimated independently, and survival of all wolves estimated jointly. The solid symbols represent Bayesian estimates using OpenBUGS, the open symbols represent maximum likelihood estimates using the kfdnm package, and the gray symbol represents the known-fate survival estimate (not including emigration) from Adams et al. (2008). (B) Estimated mean

number of wolves added to each pack in May of each year from 1987-1990 assuming survival differed between groups. Gray symbols represent observed mean number of pups per pack on October 1 from Adams et al. (2008). The dark error bars represent 95% credible intervals for the Bayesian estimates and 95% confidence intervals for the maximum likelihood estimates.

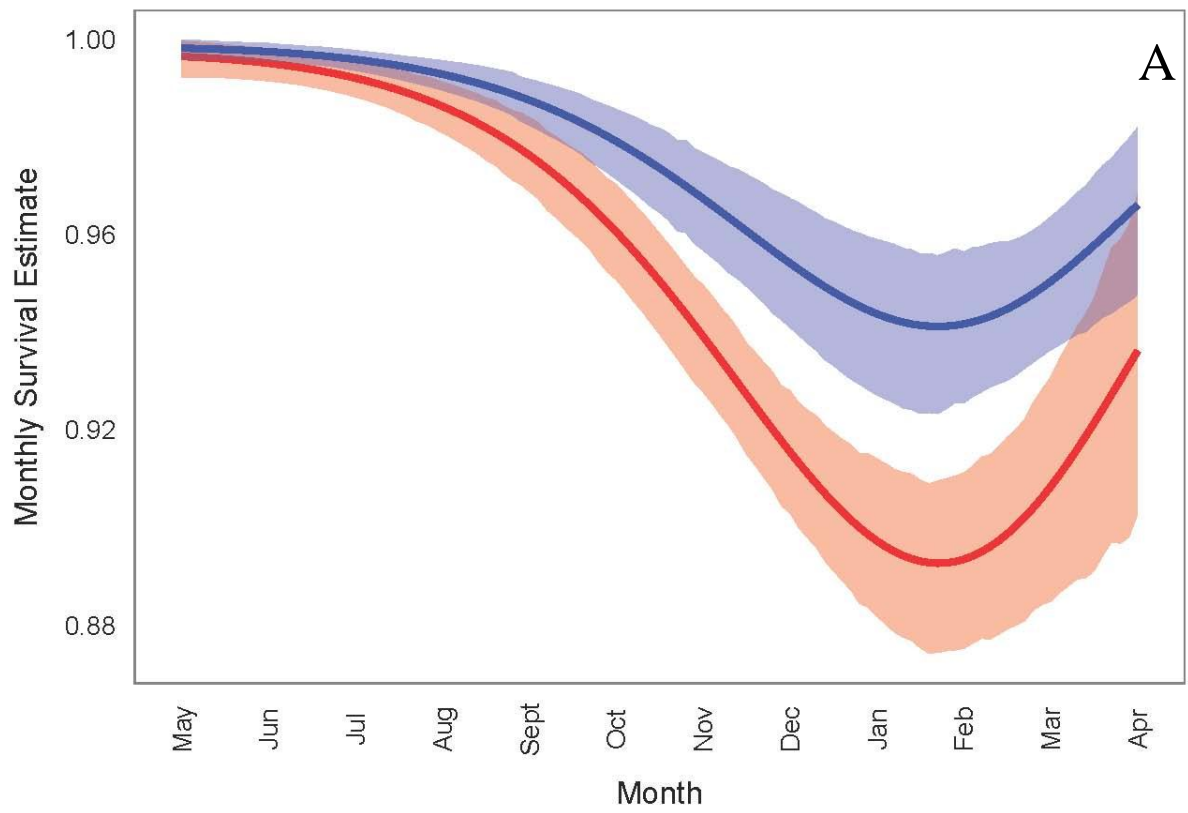

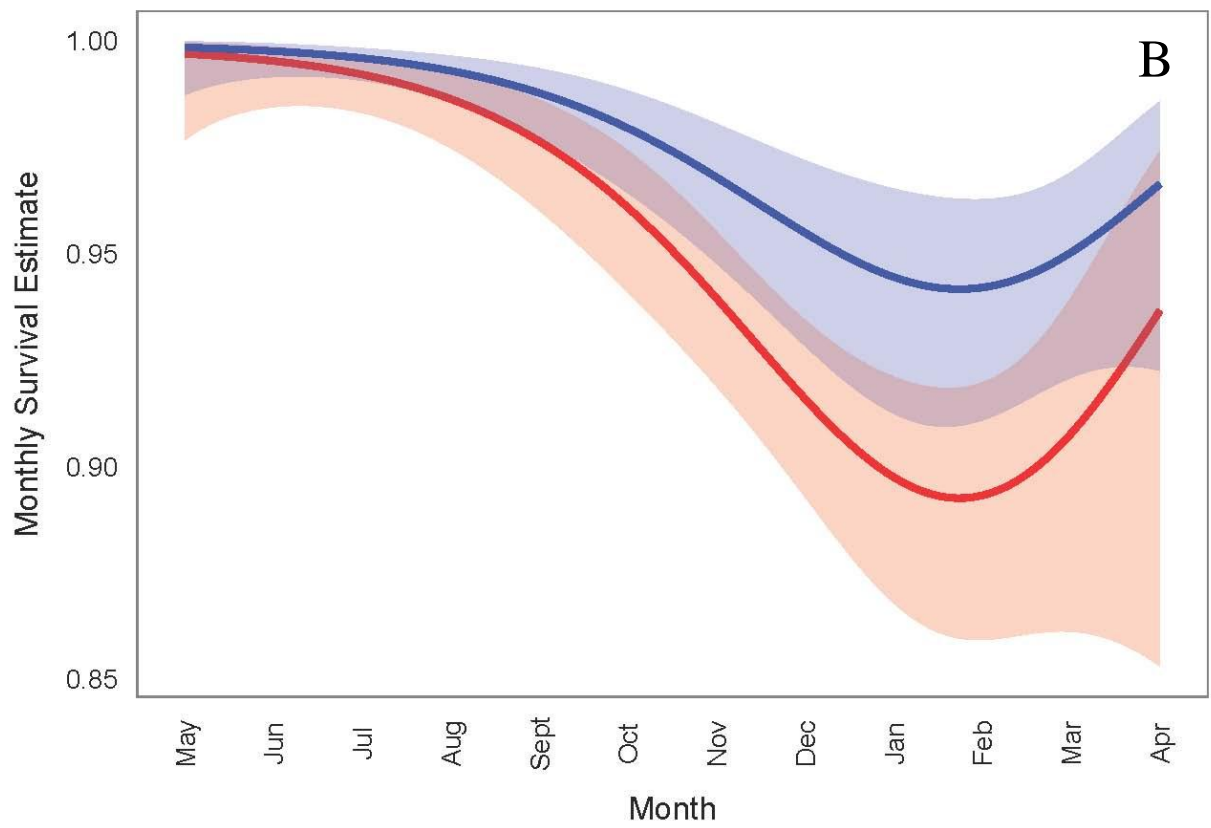

Fig. 2. Estimated average monthly survival rate for collared (blue) and uncollared (red) wolves in the study population in Gates of the Arctic National Park and Preserve, Alaska from 1987-1990. Results are based on a model assuming a common pattern in survival across months for both collared and uncollared wolves, with overall rates differing between the two groups. Models were fit using the kfdnm R package. Both Bayesian (A) and maximum-likelihood estimates (B) are presented to allow a comparison of estimates of precision (colored bands represent 95% intervals).

Appendix S1. OpenBUGS code for the combined known-fate open *N*-mixture model.

```
model  {

#Priors
 #Survival
 beta.int~dunif(-5,5)
 beta1.int~dunif(-5,5)

 for(i in 2:9){
  beta[i]<-beta.int
  beta1[i]<-beta1.int
  }
 #Recruitment
 for(i in 1:5){
  mean.count[i]~dunif(-3,3)
  recruit3[i]~dunif(-3,3)
  }
 #Detection
 p.int[1]~dunif(0,1)
 p.int[2]<-0.99999

 #Between-year survival (April-May)
 btw.yr.surv1<-1/(1+exp(-beta.int))
 btw.yr.surv2<-1/(1+exp(-beta1.int))

 #May-Aug survival, period 1
 beta[1]<-logit(pow(1/(1+exp(-beta.int)),3))
 beta1[1]<-logit(pow(1/(1+exp(-beta1.int)),3))

#Model for collared animals
for(i in  1:144){
  mu[i,1]<-collar.prev.yr[i]*(btw.yr.surv1)+collar.not.prev[i] #if previously collared, estimate between-year survival
  y.collar[i,1]~dbern(mu[i,1])
 for(j in (first[i]+1):last[i]){           #Basically a nest survival model here
  logit(phi[i,j-1])<-beta[j-1]              #Collared animal survival
  mu[i,j]<-phi[i,j-1]*y.collar[i,j-1]
  y.collar[i,j]~dbern(mu[i,j])
  }
  }

#Model for counts of rest of pack members
#Getting initial counts in May (beginning of year)
  for(i in 1:16){                           #Study initiated, 16 packs marked
   n[i,1]~dpois(mean.count1[i,1])           #Initial number in group
   log(mean.count1[i,1])<-mean.count[year.count[i]]    #Year-specific initial group size
```

```
    n1[i,1]<-n[i,1]                            #May only
      }
   for(i in 17:25){                            #9 newly marked packs, some collars added
    n[i,1]~dpois(mean.count1[i,1])             #Initial number in group
    log(mean.count1[i,1])<-mean.count[year.count[i]]    #Year-specific initial group size
    n1[i,1]<-n[i,1]-new.col[i,1]               #Accounting for removals to the collared sample
      }

   for(i in 26:88){                            #packs marked in previous years
     n[i,1]~dbin(btw.yr.surv2,n1[prev.count.pos[i],10])  #Between year survival (i.e. Apr-May)
     n1[i,1]<-n[i,1]+(recruit[i])-new.col[i,1]          #Survived, recruits, remove new collars
      }
#Recruitment submodel (recruits only added in May (period 1))
for(i in 1:88){
  recruit[i]~dpois(recruit2[i])
  lrecruit2[i]<-recruit3[year.count[i]]
  recruit2[i]<-exp(lrecruit2[i])
   for(j in 1:10){                             #10 revisit periods, 9 intervals (May-Aug = inteval 1)
    pp[i,j]<-p.int[test[i,j]]                  #Test indicates whether counts were perfect
    y.count[i,j]~dbin(pp[i,j],n1[i,j])
     }
#Survival of uncollared animals
   for(j in 2:10){
    lphi.1[i,j-1]<-beta1[j-1]
    phi.1[i,j-1]<-1/(1+exp(-lphi.1[i,j-1]))
    n1[i,j]<-n[i,j]-new.col[i,j]               #Remove newly collared individuals from the count
    n[i,j]~dbin(phi.1[i,j-1],n1[i,j-1])
     }
    }
 }
```

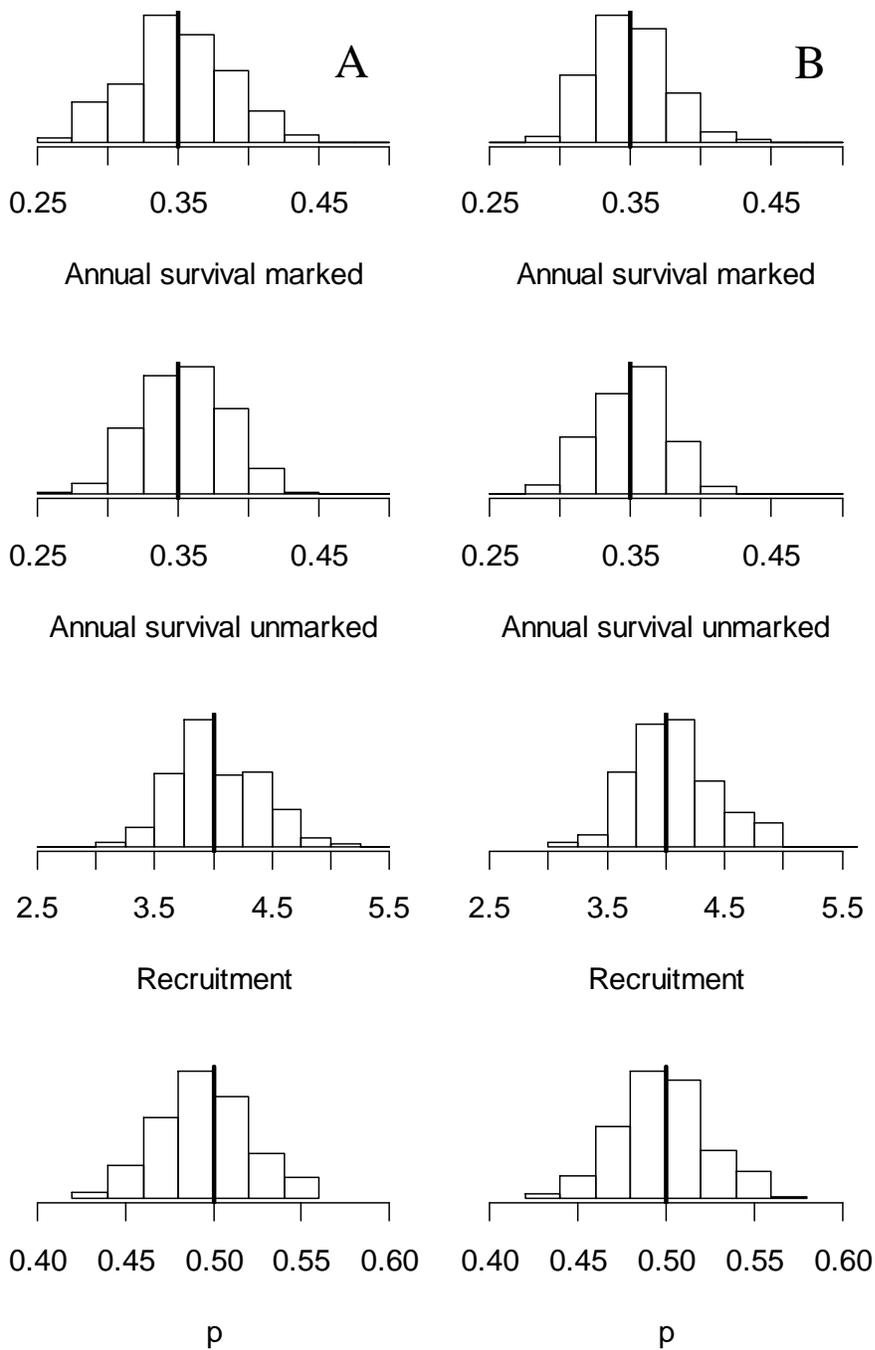

Fig. S2.1. Summaries of the estimated mean parameter values from a model fitted in OpenBUGS (A) and kfdnm (B) to 200 simulated data sets where survival is estimated separately for marked and unmarked samples. True values are indicated by the black vertical line.

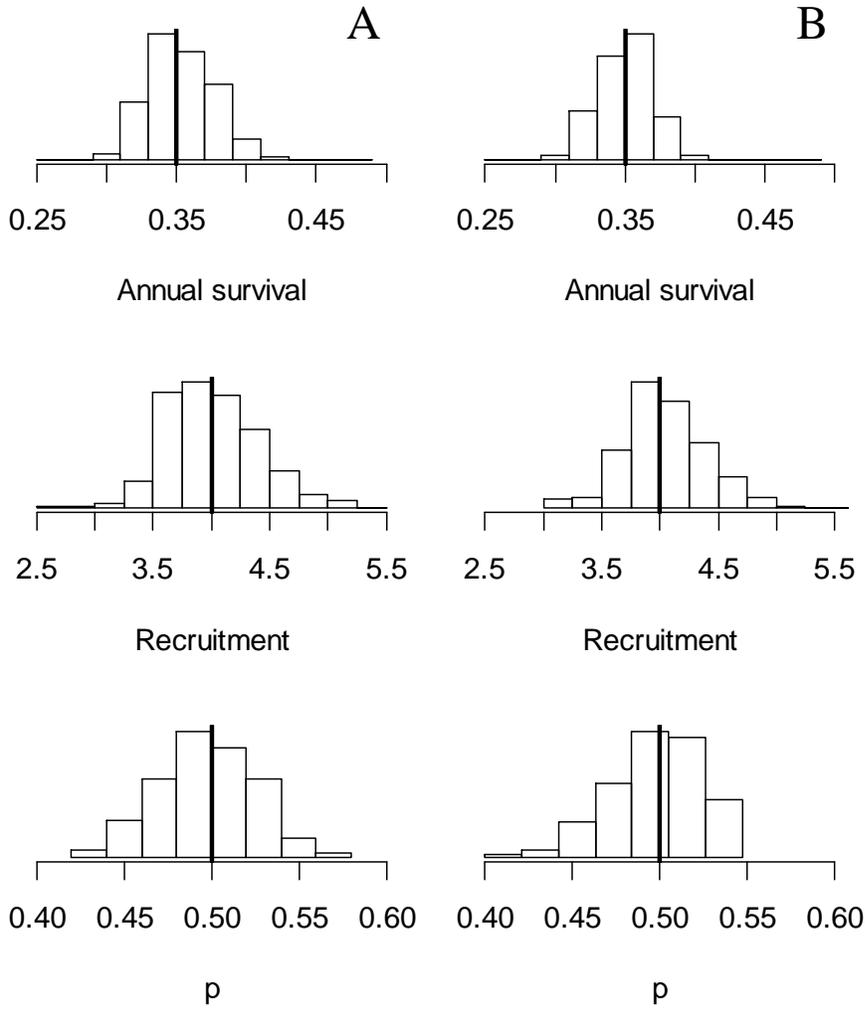

Fig. S2.2. Summaries of the estimated mean parameter values from a model fitted in OpenBUGS (A) and kfdnm (B) to 200 simulated data sets where survival is the same for marked and unmarked individuals and is jointly estimated. True values are indicated by the black vertical line.